\documentclass[useAMS,usenatbib]{mn2e}

\usepackage{graphicx}
\usepackage{amssymb}
\usepackage{subfigure}

\def\trelax{{t_{\rm r}}} 
\def\trelaxzero{{t_{\rm r,0}}} 

\title{The Tidal Tails of the Ultra-Faint Globular Cluster Palomar 1}
\author[M. Niederste-Ostholt et al.]
{M.Niederste-Ostholt$^1$\thanks{E-mail:mno@ast.cam.ac.uk}, V. Belokurov$^1$,
  N.W. Evans$^1$, S. Koposov$^{1,2}$,
  M. Gieles$^1$,\newauthor
  M.J. Irwin$^1$\\
  $^1$ Institute of Astronomy, Madingley Rd, Cambridge, CB3 0HA\\
  $^2$ Sternberg Astronomical Institute, Universitetskiy pr. 13, 
  119992 Moscow, Russia}
\begin{document}

\date{May 2010}

\voffset-.6in

\pagerange{\pageref{firstpage}--\pageref{lastpage}} \pubyear{2010}

\maketitle

\label{firstpage}
\vspace{-20pt}

\begin{abstract}
  Using the Optimal Filter Technique applied to Sloan Digital Sky
  Survey photometry, we have found extended tails stretching about
  $1^\circ$ (or several tens of half-light radii) from either side of
  the ultra-faint globular cluster Palomar 1. The tails contain
  roughly as many stars as does the cluster itself.  Using deeper
  Hubble Space Telescope data, we see that the isophotes twist in a
  chacteristic S-shape on moving outwards from the cluster centre to
  the tails. We argue that the main mechanism forming the tails may be
  relaxation driven evaporation and that Pal 1 may have been accreted
  from a now disrupted dwarf galaxy $\sim 500$ Myr ago.
 \end{abstract}


\section{Introduction}

\begin{figure}
	\centering
	\includegraphics[width=0.4 \textwidth]{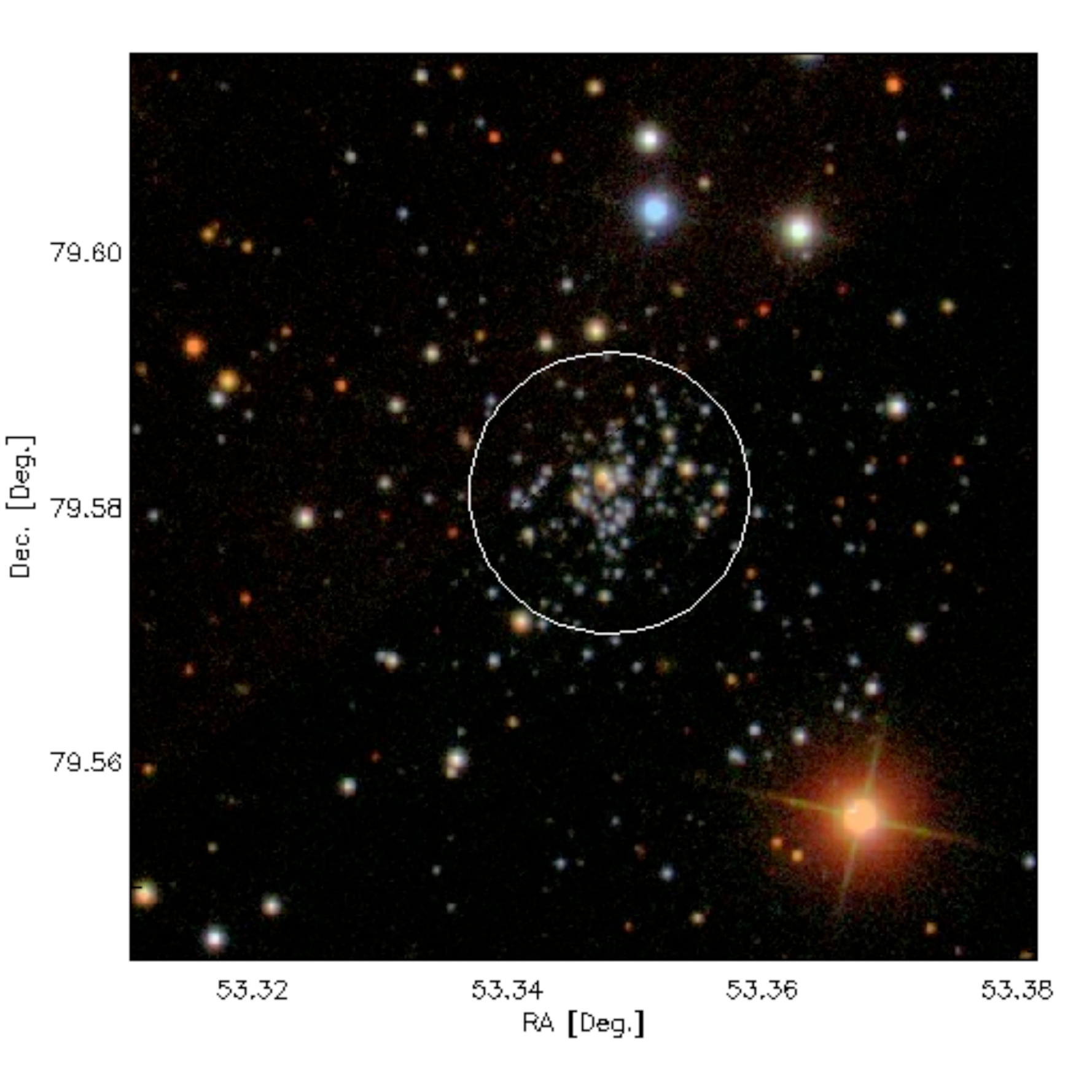}
	\includegraphics[width=0.4 \textwidth]{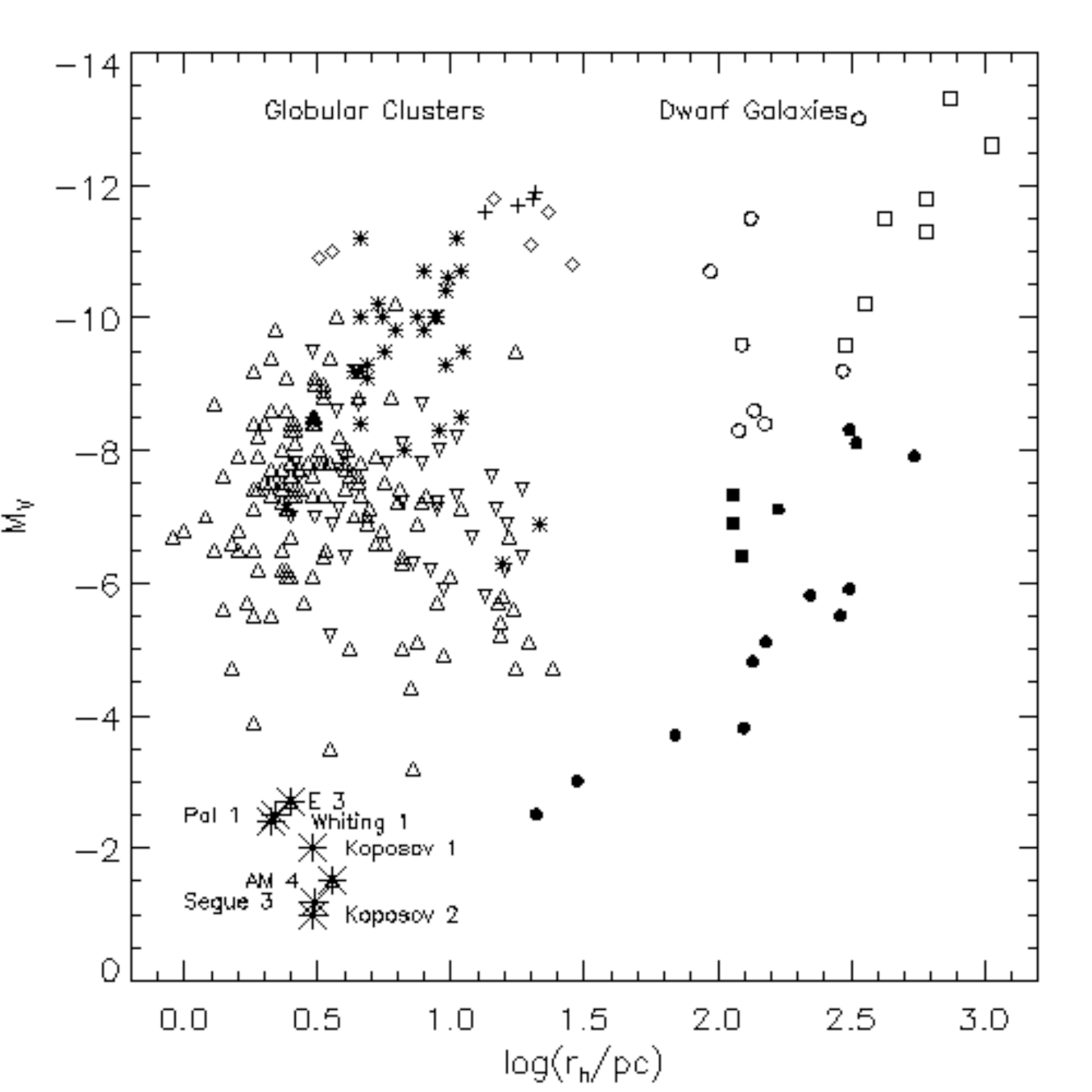}
	\vspace{-8pt}
        \caption[SDSS Image of Pal 1]{Upper: SDSS image of Pal 1. The
          white circle marks the half-light radius of the cluster
          found by~\citet{Ro98a}. Lower: Absolute magnitude versus
          half-light radius for various globular clusters and dwarf
          galaxies. Pal 1 is among a group of ultra-faint globular
          clusters in the lower left corner.}
	\label{fig:pal1_image}
\end{figure}

Fig.~\ref{fig:pal1_image} shows a Sloan Digital Sky Survey (SDSS)
image of the sparsely populated, young halo globular cluster Palomar
1. It was originally discovered by \citet{Ab55}, and lies $3.7$ kpc
above the Galactic disk and $17.3$ kpc from the Galactic Centre
\citep{Ro98a,Ro98b,Sa07}. Its size and low luminosity are very similar
to recently discovered Milky Way globular clusters, such as Segue 3,
Koposov 1 and 2 \citep{Be10,Ko07}, as well as Whiting 1, AM 4 and E
3~\citep{Wh02,Ca05,vdB80, Ca09}. The Sloan Digital Sky Survey color-magnitude
diagram (CMD) of Pal 1 shows a red clump, a main sequence turn-off and
a well defined main-sequence down approximately two magnitudes from
the turn-off. The giant branch on the other hand is very sparsely
populated \citep{BS95,Sa07}.

Owing to its unusually flat mass function, \citet{Ro98a} suggested
that Pal 1 has either experienced strong dynamical evolution (either
tidal shocks or evaporation) or that its initial mass function is
significantly different from other halo clusters. \citet{vdB00} notes
that Pal 1 is part of a group of young halo globular clusters
(including Palomar 12, Ruprecht 106, IC 4499, Arp 2, Terzan 7, Palomar
3, Palomar 4, Eridanus, Fornax 4, NGC 4590). It has often been
suggested that these young globular clusters were formed in dwarf
companions of the Milky Way, which have since been accreted and
destroyed, leaving their clusters behind. For example, \citet{Cr03}
proposed that Pal 1 is part of the Monoceros ring (along with NGC
2808, NGC 5286, NGC 2298), whilst \citet{Be07} suggested it may be
associated with the accretion event that formed the Orphan Stream
(along with Ruprecht 106 and possibly Terzan 7).

Pal 1, together with Segue 3, Koposov 1 and 2, Whiting 1, E 3, and AM
4, comprise {\it an ultra-faint population} of globular clusters. Pal
1 is the nearest of the six contained within the SDSS footprint
(though E3 is still closer to the Sun and Galactic Centre). Compared
to the other young globular clusters, Pal 1 is the faintest and
smallest. Its tiny size and low luminosity (and hence presumably low
mass) make it an attractive target to look for the effect of tides or
dynamical evolution, which is the purpose of this {\it Letter}.

\vspace{-10pt}
\begin{figure}
	\centering
	\includegraphics[width=.5 \textwidth]{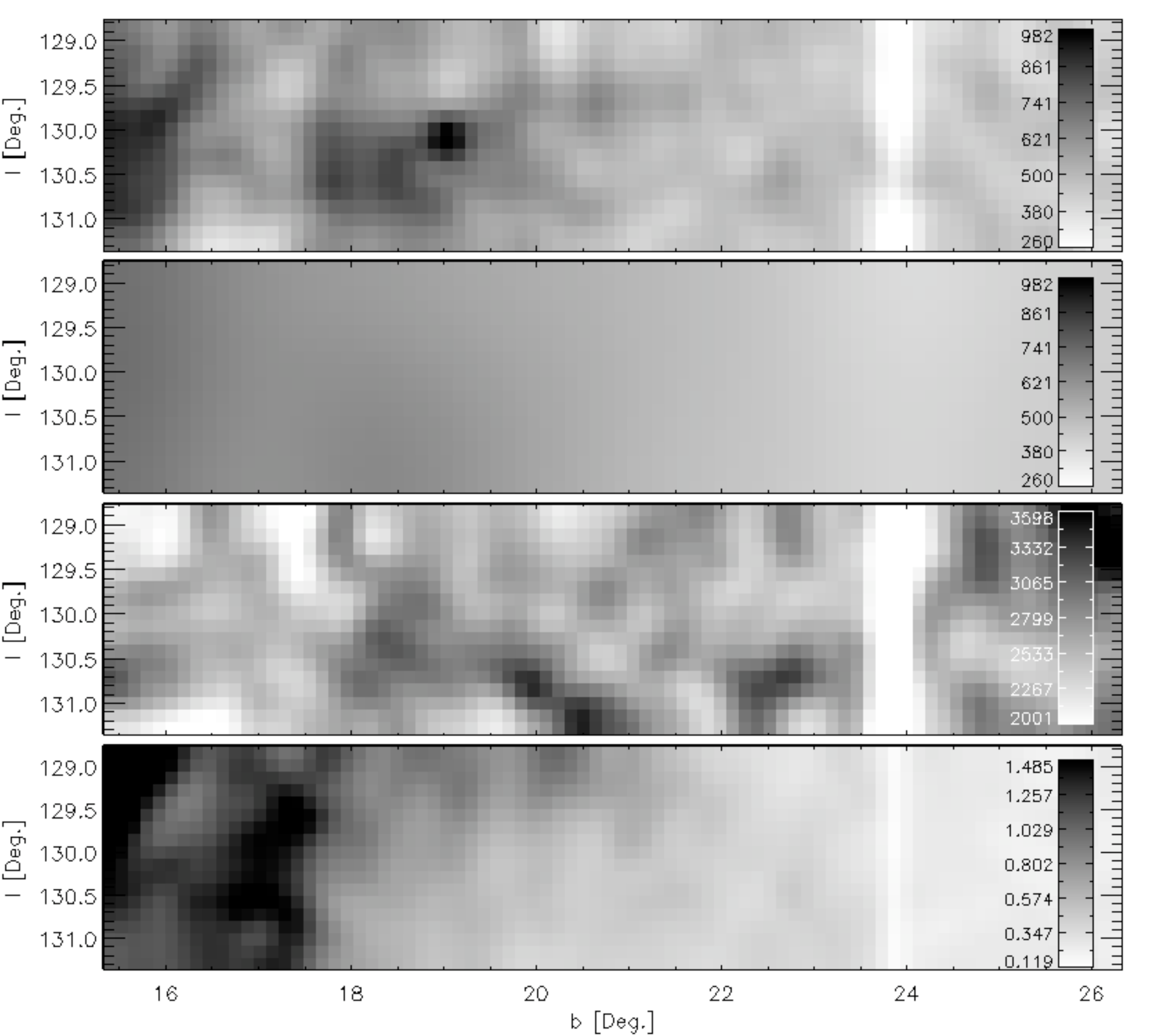}
	\vspace{-8pt}
        \caption[SDSS Coverage of Pal 1]{Upper: The density of stars
          around Pal 1 lying within a CMD mask drawn around the
          cluster's distribution (see Fig.~\ref{fig:cmfollowup_pal1} left panel). The
          conversion of grey-scale to stars per degree is given in the
          bar on the right. Upper middle: An estimate of the smooth
          field star population around Pal 1 using the same
          grey-scale. Lower middle: The density of galaxies around Pal
          1. Lower: The mean $g$-band extinction per pixel around Pal
          1.}
	\label{fig:pal1_overview}
\end{figure}

\section{Tail Detection}

Pal 1 is located at $(l,b)=(130.1^{\circ},19.0^{\circ})$. It has an
absolute magnitude of $M_V=-2.5\pm0.5$ and a half-light radius of
$R_{\rm h}\approx2.2$ pc and its heliocentric radial velocity is
$-82.3\pm3.3$ kms$^{-1}$ \citep{Ro98a,Ro98b,OR85}. Pal 1 lies within
an SDSS data release 7 stripe~\citep{Ab09}, covering $2.5^{\circ}$ in
galactic longitude and $11.3^{\circ}$ in galactic latitude. Henceforth,
all SDSS magnitudes are corrected for extinction using the maps of
\citet{Sc98}.

The panels of Fig.  \ref{fig:pal1_overview} show $82\times 18$ pixel
maps of the the density of stars around Pal 1, an estimate of the
possible smooth underlying distribution of stars, the distribution of
galaxies in the area, as well as the mean $g$-band extinction.  Pal 1
is clearly visible as an overdensity in the topmost panel. To search
for stars torn from Pal 1, we employ the optimal filter technique
which works by calculating conditional probabilities of cluster and
foreground membership from densities in colour-magnitude space, known
as Hess diagrams. From this weighted distribution, a smooth
distribution of field stars is then subtracted \citep[see
e.g.][]{Od03,Ni09}. The field star density is estimated by removing a
$0.3^{\circ}\times0.3^{\circ}$ box around the cluster, replacing it by
a patch at $(\ell,b)=(131.1^{\circ},22.0^{\circ})$, and smoothing the
resultant distribution with a Gaussian kernel of FWHM $15$ pixels and
a box-car smoothing over $2$ pixels. 

\begin{figure*}
	\centering
	\includegraphics[width=0.3 \textwidth]{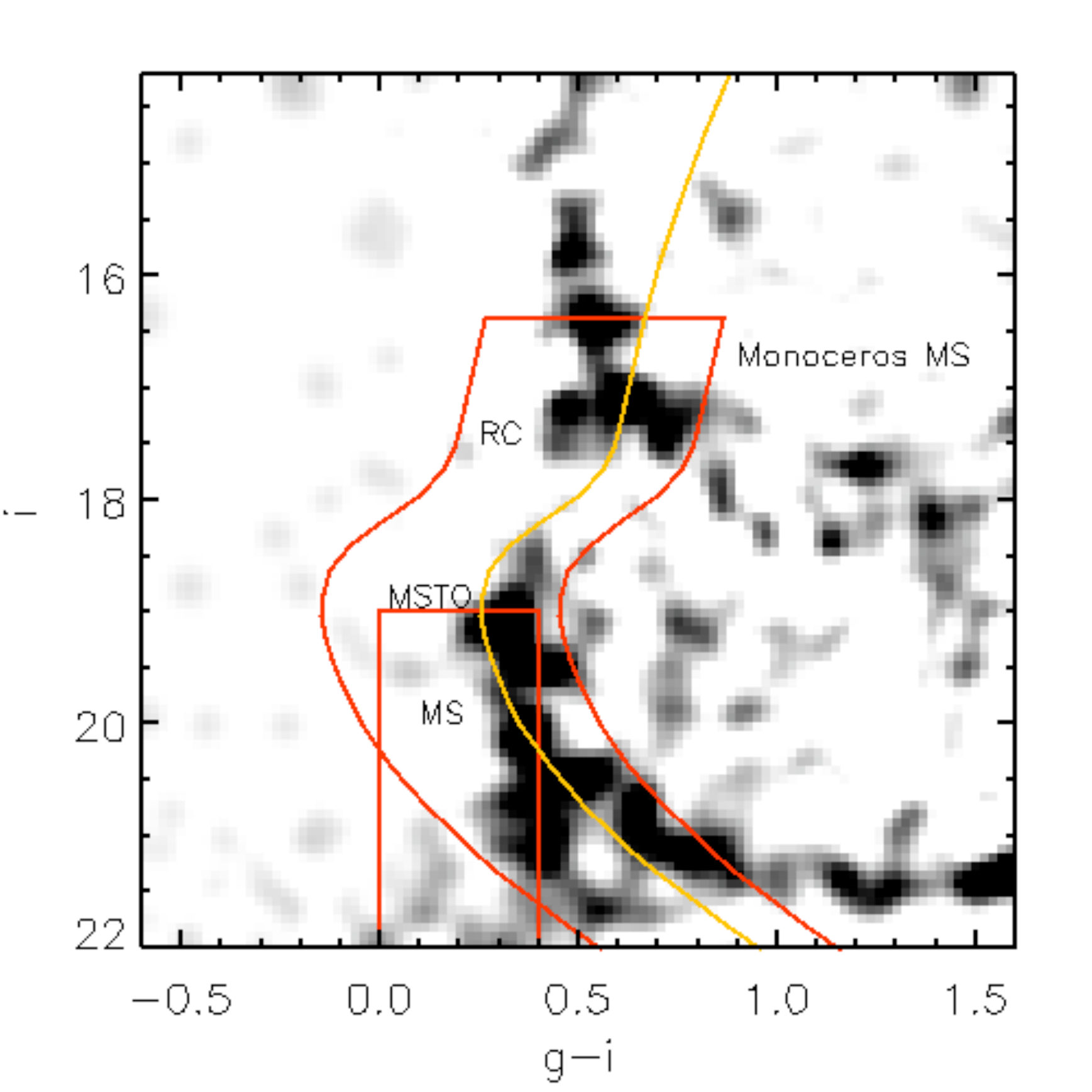}
	\includegraphics[width=0.3 \textwidth]{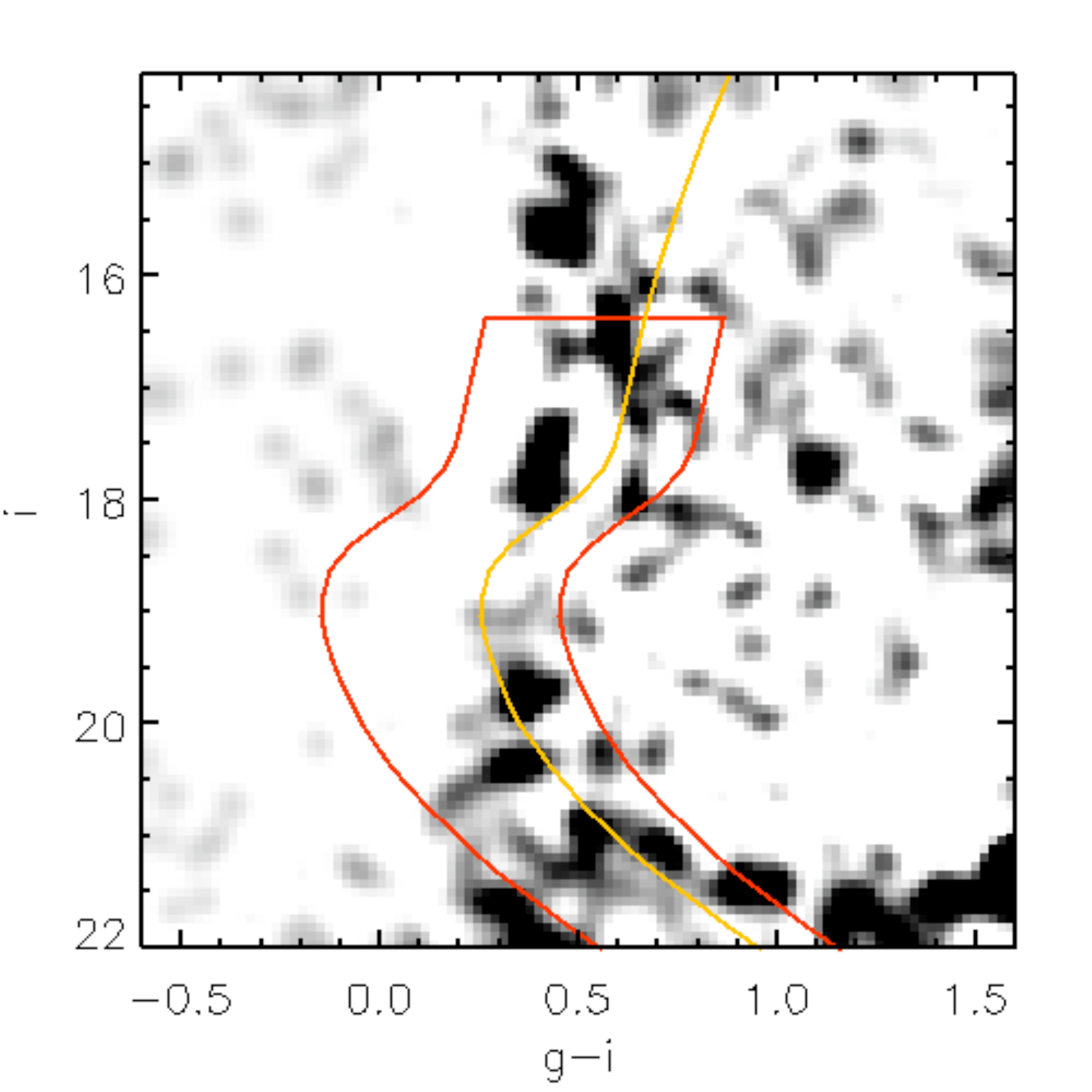}
	\includegraphics[width=0.3 \textwidth]{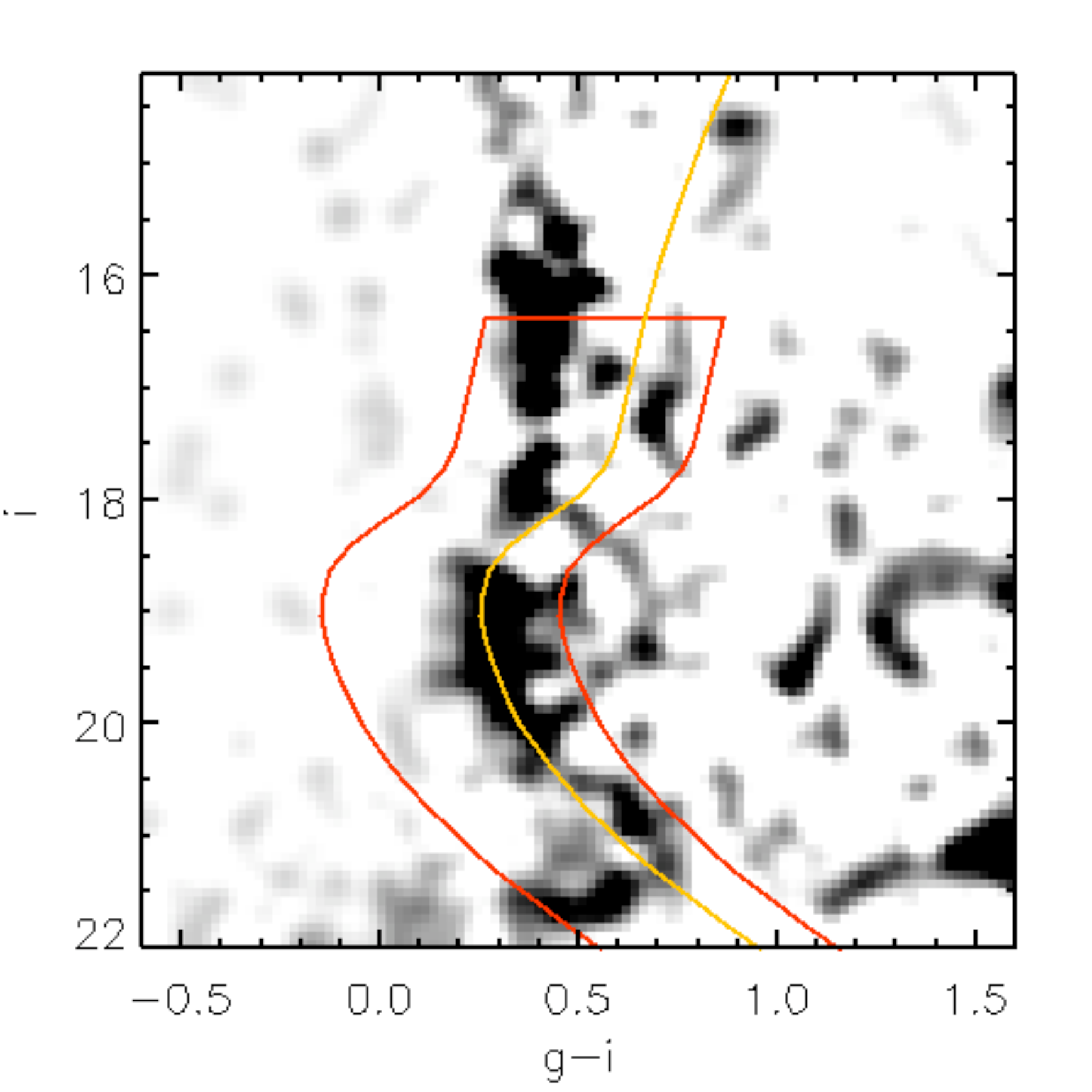}
	\vspace{-8pt}
        \caption[Hess Difference Diagrams around of Pal
        1]{Left: Hess difference
         between stars within $0.16^{\circ}$ from the centre of Pal 1
          and stars further than $0.5^{\circ}$ from the centre (field
          stars). The mask is used to select stars for the optimal
          filter analysis, with stars outside the mask having zero
          weight. Also shown is the rectangular box mask used in the
          star-count analysis. The ridgeline overplotted for
          comparison purposes is that of the metal-poor globular
          cluster M92 \citep{Cl05}, offset to Pal 1's distance
          modulus. Pal 1's main sequence (MS), main sequence turn-off
          (MSTO), and red clump (RC) are labelled. The structure
          visible approximately $3$ magnitudes brighter than Pal 1 is
          probably the main sequence of the Monoceros ring.
          Center: The Hess difference diagram of stars in the northern (higher latitude)
          tail of debris around Pal 1 together with the CMD mask used
          to select cluster stars. Right: The same, but for the the
          southern (lower latitude) tail. In both cases, the Hess diagram
          of field stars in the area (defined as stars
          with $b>21.5^{\circ}$ and $b<17.5^{\circ}$) is
          subtracted. Both
          stream candidates show an overdensity of stars near Pal 1's
          main-sequence turn-off with a slight extension down the main
          sequence. In all cases, the ridgeline of the metal-poor
          globular cluster M92 \citep{Cl05} is overplotted in yellow.}
	\label{fig:cmfollowup_pal1}
\end{figure*}

The left-hand panel of Fig.~\ref{fig:cmfollowup_pal1} shows the Hess difference between stars
within $0.16^{\circ}$ and those further than $0.5^{\circ}$ from the
centre of Pal 1. The mask shown encloses both the main sequence and
the red clump, though the inclusion of the red clump makes little
difference to our results as the giant-branch is so poorly
populated. The mask closely follows the main sequence at the red edge
but allows more leeway on the blue side. This is done since the
contamination from disk and halo stars is more severe at the red side
of the distribution. The ratio of the Hess diagrams is used as the
weights in the optimal filter analysis.


\begin{figure*}
        \centering
        \includegraphics[width=0.9 \textwidth]{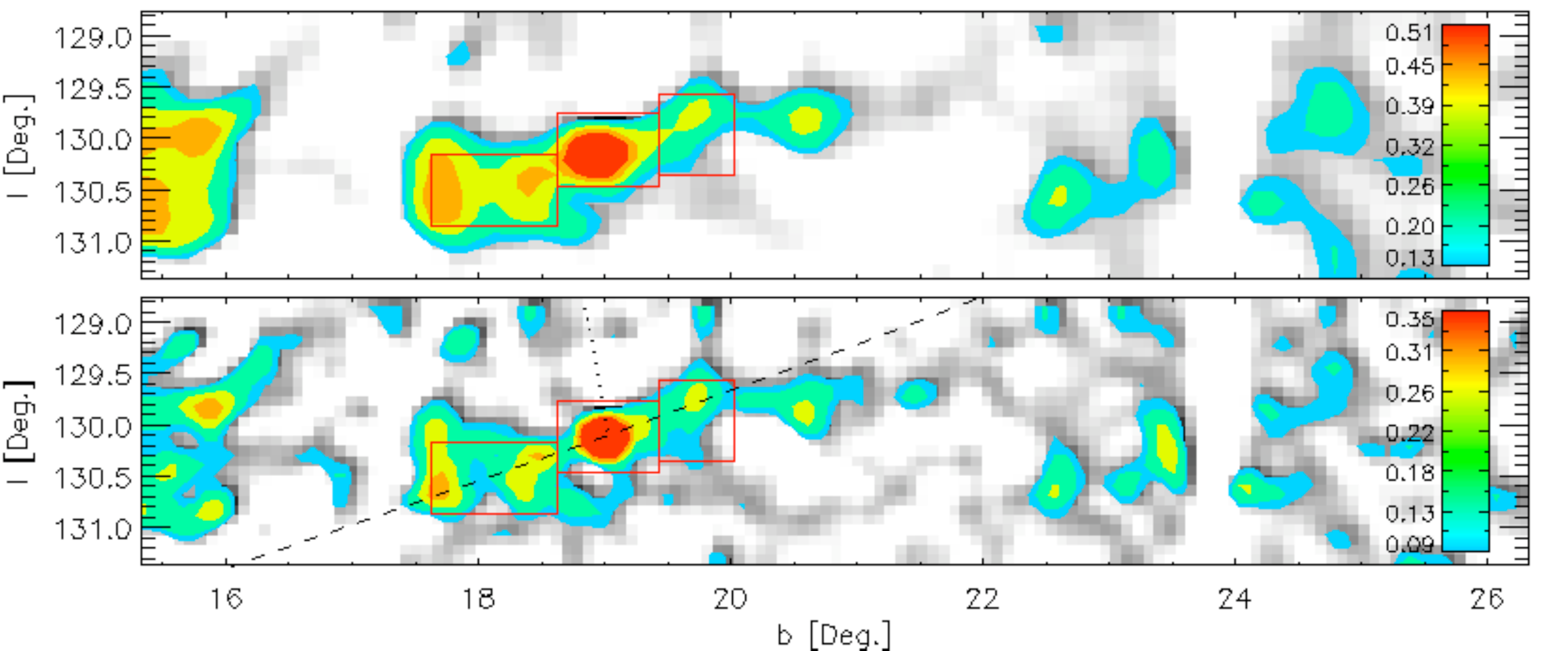}
        \vspace{-8pt}
        \caption[Optimal Filter Analysis of Pal 1]{Upper: Optimal
          filter technique applied to SDSS data centred on Pal 1. The
          area is divided into $82\times18$ pixels and smoothed with a
          Gaussian kernel of $FWHM=3$ pixels. The coloured contours
          show $1.5,2,3,4,5 \sigma$ levels above the mean. The density
          is normalized to the average pixel density within
          $0.^{\circ}08$ of the centre of Pal 1. The colour within the
          contours can be converted into normalized pixel density by
          the key.  The cluster clearly stands out in the centre of
          the plot. There are two overdensities connected to Pal 1,
          extending away from it. Aside from a prominent overdensity
          at lower latitudes, the field is quite smooth. The boxes
          mark the areas contain probable debris of Pal 1 and whose
          Hess difference diagrams are shown in
          Fig.~\ref{fig:cmfollowup_pal1}. Bottom: Optimal filter
          analysis with a higher spatial resolution ($123\times27$
          pixels). The structural integrity of the two extensions from
          Pal 1 is preserved and they appear to be streams showing a
          characteristic S-shape near the cluster. The dashed line is
          a great circle fitted to the possible debris structure with
          a pole at $(\ell,b)=(-132.2^{\circ},21.3^{\circ})$. The
          dotted line shows the direction towards the Galactic
          centre.}
        \label{fig:pal1_optfilter}
\end{figure*}

Fig.~\ref{fig:pal1_optfilter} shows the results of the optimal filter
analysis. The distribution is smoothed with a Gaussian kernel over $3$
pixels. Pal 1 stands out as the most significant overdensity in the
field but there are significant overdensities extending from it
towards higher and lower galactic latitudes. There is also a high
signal region at the lower edge of the plot, but follow-up analysis
shows this to be unrelated to the cluster and likely to be an artefact
caused by the high extinction in this region, together with edge
effects. The lower panel of Fig.~\ref{fig:pal1_optfilter} shows the
results with a higher spatial resolution ($123\times27$ pixels),
smoothed over $3$ pixels. Again, the tails are noticeable in an
otherwise nearly noise-free area. Color-magnitude analyses of the
tails are shown in Fig.~\ref{fig:cmfollowup_pal1}. There are
suggestive similarities between the main body of Pal 1 and this
debris. The stars in both extensions are concentrated around Pal 1's
main sequence turn-off with a faint continuation down the main
sequence.

Pal 1 does not appear to be embedded in a noisy field. It is not
possible to conclude from the optimal filter analysis whether or not
it was once part of a dwarf companion. Of course, such a stream may be
hidden by the small field-of-view available, or it may already have
dispersed.

\begin{figure}
	\centering
	\includegraphics[width=0.49 \textwidth]{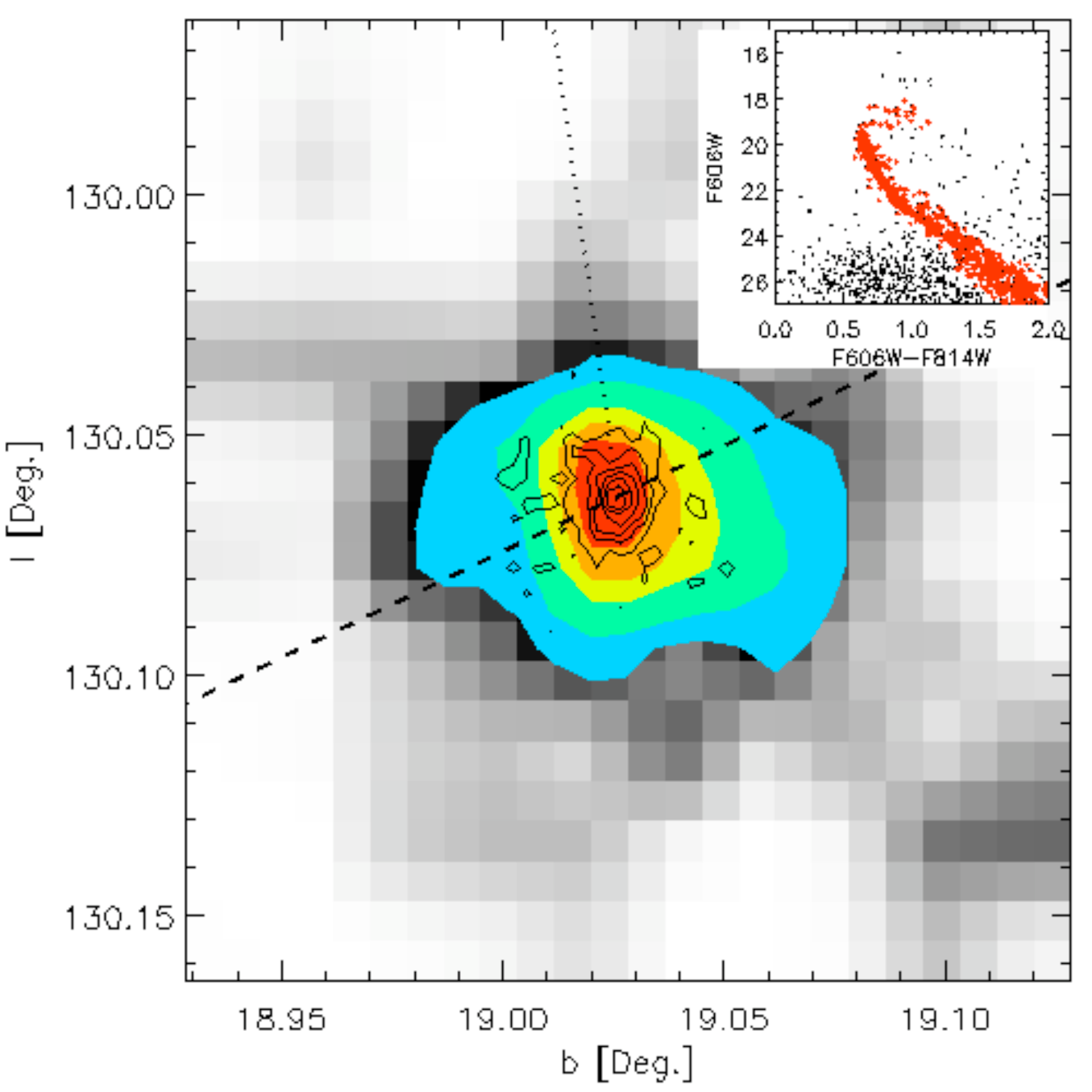}
	\vspace{-8pt}
        \caption[The Centre of Pal 1]{Inset: CMD of the HST
          observations of Pal 1 in the F606W and F814W filters
          (approximately $V$ and $I$). The red dots mark the stars
          which are included in the black-line contours. Main: High
          resolution optimal filter analysis (grey-scale and colored
          contours) with HST contours overplotted. The coloured
          contours show the $0.5, 1.5, 3, 4, 5 \sigma$ levels above
          mean density in the field. This corresponds to densities (in
          number of stars per square arc minute) of 78.1, 156.3,
          234.4, 312.5, 390.6, 468.8.  The central region of the
          cluster is elongated and points towards the Galactic Centre
          (dotted line). The dashed line shows the direction of the
          debris.}
	\label{fig:opt_hst}
\end{figure}

There exist deep {\it Hubble Space Telescope} (HST) ACS data covering
Pal 1~\citep{Sa07}, which allow us to explore the central regions more
closely. By applying our source extraction to the HST images, we
select cluster stars based on their position in color-magnitude space,
as shown in the inset in Fig.~\ref{fig:opt_hst}.  We then overplot the
density contours derived from the HST data on a higher resolution
($574$ by $126$ pixels, smoothed over $1$ pixel) version of the
optimal filter analysis. This high resolution plot of the central
$0.2^{\circ}\times0.2^{\circ}$ does not allow us to identify any
debris. However, we can see that the central region of the cluster is
elongated, pointing towards the Galactic Centre, and that the cluster
stars are spread over several half-light radii ($R_{\rm h} \approx
0.01^{\circ}$). The outer isodensity contours, given by the optimal
filter technique, twist and align with the direction of the
debris. Such an S-shaped misalignment between tidal debris and the
elongation of the central regions is observed in simulations when the
object is near the apocentre of its orbit \citep[see for example the
bottom row of Fig.~4 in ][]{Jo02}.

\begin{figure}
	\centering
	\includegraphics[width=0.5 \textwidth]{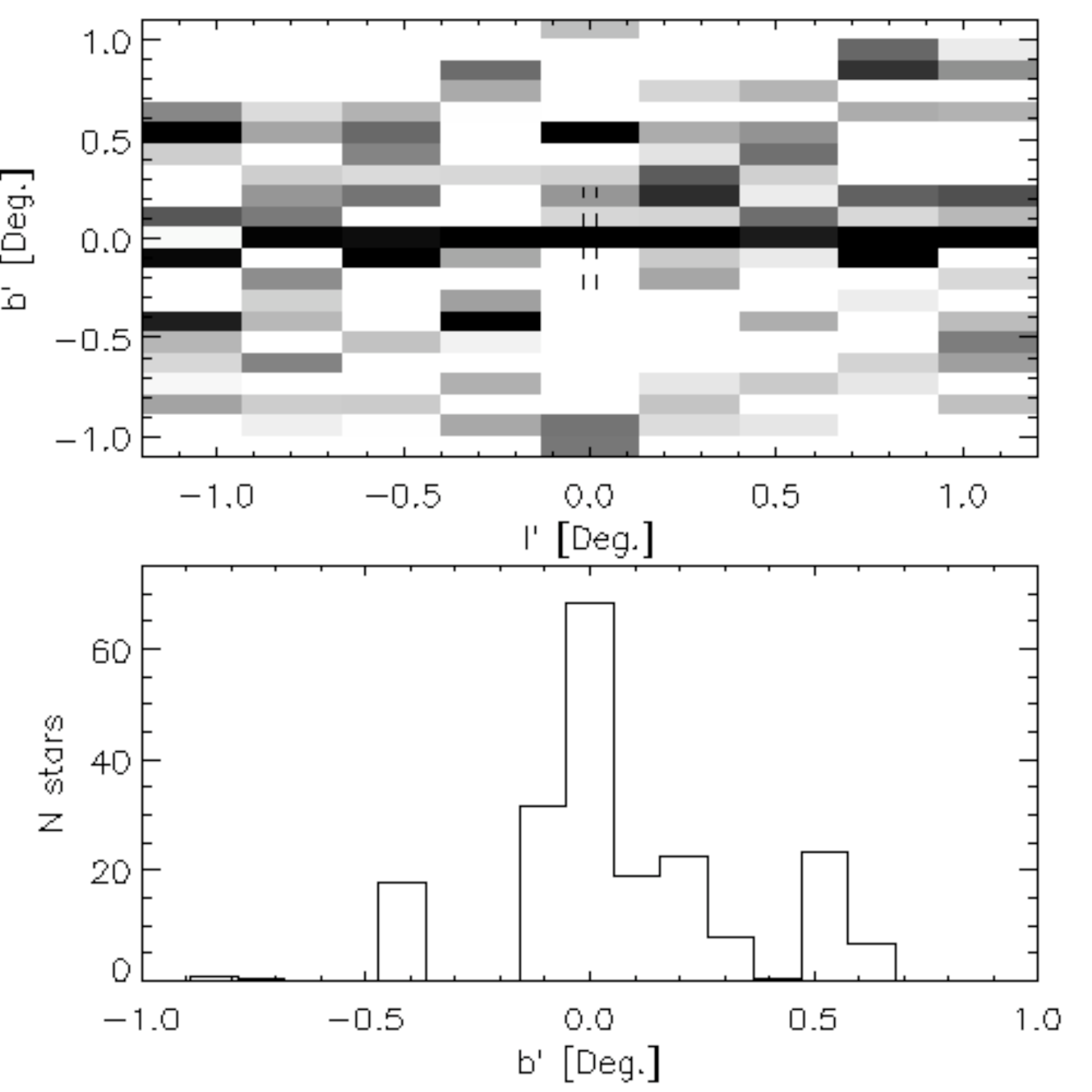}
	\vspace{-8pt}
        \caption[Pal 1 Star Counts]{Top: Background subtracted density
          of stars in the $(\ell',b')$ coordinate system, which is
          defined by its pole at $(\ell,b) =
          (-132.2^\circ,21.3^\circ)$ and origin at Pal 1's
          location. The core of the cluster is shown as two vertical
          lines. We identify the continuous high density region lying
          along $b' \approx 0^{\circ}$ as the Pal 1 streams. Bottom:
          Number of stars vs. latitude. There is a significant spike
          at $b'\approx 0^{\circ}$ suggesting there are approximately
          $60$ stars in the debris, comparable to the number of stars
          within Pal 1.}
	\label{fig:starcounts}
\end{figure}

To estimate the number of stars in the tails, we use a coordinate
system ($\ell',b'$), in which the extensions lie along a great circle
(see lower panel of Fig.~\ref{fig:pal1_optfilter}). In this system,
Pal 1's possible debris lie at roughly constant latitude ($b'\approx
0^{\circ}$).  The top panel of Fig.~\ref{fig:starcounts} shows the
density of stars in a $2^\circ$ area around Pal 1 in this new
coordinate system. There are $9$ bins along, and 21 bins perpendicular
to, the stream direction.  By using larger bins in the direction along
the stream, we can assess the significance of the enhancement around
Pal 1.  The influence of the cluster itself has been removed from this
plot by excluding stars that are within $4$ half-light radii ($\sim
0.04^{\circ}$) of the cluster centre. To amplify the signal of Pal 1
stars, this plot is restricted to stars lying in the rectangular box
in color-magnitude space shown in Fig.~\ref{fig:cmfollowup_pal1} left panel, which
picks up the Pal 1 main sequence turn-off and the blue edge of the
main sequence. This avoids contamination from the field stars, which
lie slightly to the red of the main sequence. To account for possible
contamination in this plot, we subtract a background estimate, derived
by applying a linear fit to each column of constant $\ell'$, to arrive
at the upper panel of Fig.~\ref{fig:starcounts}. The stream stands out
as a continuous overdensity of pixels along $b' \approx 0$. It is now
straightforward to count the number of stream stars present by summing
the weighted, background subtracted counts in bins of constant
latitude $b'$ as shown in the lower panel of
Fig.~\ref{fig:starcounts}. There is a significant spike at $b'\approx
0^{\circ}$, which we identify as the tails of Pal 1. There seem to be
on order the same number of stars in the tails (approximately $70$) as
in the cluster (approximately $85$), suggesting that the mass in the
tails is comparable to the mass in the cluster.

\section{Conclusions}

We have discovered probable tails around the globular cluster Pal 1 in
Sloan Digital Sky Survey data. The tails cover at least $2^{\circ}$ on
the sky and extend northward and southward from the cluster centre,
possibly up to $\approx 80$ half-light radii. The tails contain
roughly as many stars as does the cluster itself. {\it Hubble Space
  Telescope} data reveal that the central parts of the cluster are
clearly elongated, with cluster stars spread over several half-light
radii. The tails constrain the direction of Pal 1's projected proper
motion, but this, together with its measured radial velocity, is not
sufficient to establish meaningful constraints on the orbital
parameters of Pal 1.

Pal 1 is a prototype of the population of ultrafaint globular clusters
with $M_v$ fainter than $-3$. This also includes Segue 3, Koposov 1
and 2, as well as Whiting 1, E3 and AM 4. If the ultrafaint globular
clusters are born in their present state, then they would evaporate
within 1 Gyr~\citep{Ro98a,Ko07}. Hence, it would be unlikely that they
survived long enough for us to see them, unless a large number of such
clusters are born like this continuously. It is much more likely that
globular clusters evolve into the ultrafaint r\'egime, either through
tides or through evaporation causing catastrophic loss of stars.  Mass
loss due to evaporation is accelerated when the cluster orbits in a
tidal field, as for example is the case if an accreted dwarf galaxy
originally hosted the cluster.

Could the tails of Pal 1 be caused primarily by evaporation?  First,
assuming a mass-to-light ratio of $\sim 2$ gives a present cluster
mass of $\sim 1400$ $M_{\odot}$. Pal 1 has a half-mass relaxation
time~\citep{Sp71} of $\trelax\sim 0.16$ Gyr.  Based on the results of
dynamical models, a lower-limit to the mass-loss rate due to
relaxation driven evaporation is 1.4 $M_\odot$ Myr$^{-1}$ for an
assumed circular orbit~\citep{Ba03}, which scales like $(1-e)^{-1}$
for an orbit with eccentricity $e$. Assuming that stars escape and
drift with velocities $\sim 1$ kms$^{-1}$ as suggested by eq. (18) of
\citep{Ku10}, then the mass within a length $L_{\rm Tail} \sim 400$ pc
of the tidal tails ($\sim 2^{\circ}$ on the sky) is $M_{\rm
  Tail}\sim\dot{M}\times L_{\rm Tail}/\sigma\sim 560 M_{\odot}$ for a
circular orbit. This rises to $\sim 1100 M_\odot$ for an eccentricity
$e=0.5$, which is comparable to the mass of the cluster. Secondly, the
Jacobi radius at Galactocentric radius of $\sim 17$ kpc for an object
of the mass of Pal 1 is $r_{\rm J} \sim 26$ pc, so that $r_{\rm
  h}/r_{\rm J} \gtrsim 0.1$.  This is surprisingly low for a
dynamically evolved system. \citet{Ba10} find that the expected
$r_{\rm h}/r_{\rm J} \sim 0.3$ for extended and low-mass Galactic
globular clusters.  This suggests that the cluster is underfilling its
Roche-surface.  Both arguments favour evaporation rather than disk
shocking as the main formation mechanism.

Pal 1 is therefore a rather different object to Pal 5, which is
well-known to have extended tails~\citep{Od03}. Pal 5 is overfilling
its Roche surface and its tails are believed to have formed primarily
by disk shocking~\citep{De04}.  Certainly, given its present state of
disruption, Pal 1 must have passed within $8$ kpc of the Galactic
Centre in order for disk shocking to account for the tidal tails, as
judged by Figure 14 of \citet{VH97}.

If Pal 1 has been evolving in a satellite galaxy with a stronger tidal
field, we expect it to have it a high density with respect to its
current Galactocentric radius.  This may explain why it is currently
underfilling its Roche surface. However, the accretion must then have
happened very recently, because the relaxation time is so low that
adjustment to the new tidal field would happen quickly.  For a 
cluster that is not too severely affected by the tidal field its (half-mass) 
relaxation time, $\trelax$, increases linearly with its age \citep{He65,Ba02} 
because of expansion driven by  binaries and 2-body relaxation. This expansion 
is ultimately stopped by the tidal field and then $\trelax$ decreases with age \citep{He61}. 
We assume that the cluster was in this Roche-lobe filling evolution in the original 
host galaxy. Let us denote by sub-script 0 
the time of accretion, and assume that Pal 1 after accretion in the Milky Way 
started expanding again due to the weaker tidal field. \citet{Gi10} showed that for 
globular clusters $ \trelax=0.3t $, where t is the age of the cluster. This means that
\begin{equation}
  \trelax \approx \trelaxzero + 0.3(t-t0) 
  \Rightarrow t-t0 \approx 3.3(\trelax-\trelaxzero)
\end{equation}
Putting $\trelaxzero\approx0$ gives a rough upper limit on the accretion time of
$\sim 500$ Myr in the past.

%


\section*{Acknowledgments} 
MNO is funded by the Gates Cambridge Trust, the Isaac Newton
Studentship fund and the Science and Technology Facilities Council
(STFC), whilst VB acknowledges financial support from the Royal
Society. We thank M.G. Walker for helpful advice and support and the referee, Laszlo Kiss, for constructive feedback.

Funding for the SDSS and SDSS-II has been provided by the Alfred P.
Sloan Foundation, the Participating Institutions, the National Science
Foundation, the U.S. Department of Energy, the National Aeronautics
and Space Administration, the Japanese Monbukagakusho, the Max Planck
Society, and the Higher Education Funding Council for England. The
SDSS Web Site is http://www.sdss.org/. The paper is partly based on
observations made with the NASA/ESA Hubble Space Telescope, obtained
from the data archive at the Space Telescope Institute. STScI is
operated by the association of Universities for Research in Astronomy,
Inc. under the NASA contract NAS 5-26555.

\label{lastpage}

\end{document}